# Dogs on forest trails; Understanding ecology of Striped Hyena and wild Canids in the presence of free-ranging dogs in Udanti-Sitanadi Tiger Reserve, Central India using Joint Distribution and Deep Neural Networks

Chiranjib Chaudhuri [1],* , Krishnendu Basak [2] , M Suraj[2] , Moiz Ahmed[2] , Amit Kumar[1]

1. AeroLine International Pvt. Ltd. : Kolkata, West-Bengal, India

Correspondance: Email: chiranjibchaudhuri@gmail.com

Email: namitkumarn@gmail.com

2. Department of Wildlife Conservation, Nova Nature Welfare Society, H. No. 36/337,

Choti Masjid, Byron Bazar, Raipur, Chhattisgarh 492001, India.

Email: bastiger08@gmail.com (Basak)

Email: mat.suraj@gmail.com (Suraj)

Email: novanature2004@gmail.com (Ahmed)

**ABSTRACT:**

This study uses Joint Species Distribution Models (JSDMs) and Deep Neural Networks (DNNs) to explore how wild carnivores and free-ranging dogs interact in the Udanti-Sitanadi Tiger Reserve (USTR) in Central India. The research focuses on key species like the Striped Hyena, Grey Wolf, Golden Jackal, and Indian Fox, revealing significant overlaps in habitat with free-ranging dogs, especially in densely populated areas like the Sitanadi region of the tiger reserve. These overlaps pose serious risks to wildlife through competition for resources, predation, and the spread of diseases. The study shows that the Striped Hyena prefers gentle slopes and forested areas, while the Grey Wolf tends to avoid cropland and thrives in regions with higher rainfall that supports a stable prey base. The Golden Jackal, more adaptable than the others, favors west-facing slopes and stable temperatures, whereas the Indian Fox is mainly found in the less disturbed, mountainous Kuladighat region. Additionally, the study highlights the potential impacts of climate change, predicting that the Grey Wolf could face habitat extinction under more severe scenarios. These findings underscore the urgent need for conservation strategies tailored to address both dog wild carnivore interactions and the growing challenges posed by climate change, focusing on protecting the critical habitats of vulnerable species like the Striped Hyena and Grey Wolf.

## Introduction

Species needs suitable habitats to thrive and to continue the evolutionary process through long term survival within the natural environment. Globally, shrinkage of these viable and crucial natural habitats is predominantly triggered by anthropogenic factors and severely affecting species survival parameters where large carnivores lost more than 90% of their distribution range (Gawrecki, 2017). Entire South-East Asia has 206.5 million ha forested land but less than 12% of these large tracts of forest areas are protected (Estoque et al, 2019). Despite having protected status, the areas are not devoid of anthropogenic impacts. The increasing human settlements in and around protected areas, increase pressure on wildlife species (Wittemyer et al., 2008), sharing the same landscape. Threats associated with human settlements are growing as well, as large population of domestic species that has various impacts on wildlife (Plaza et al., 2019).

Domestic dogs are the most abundant predator of the world (Villatoro et al., 2019) and their estimated population reached around 900 million (Gompper, 2014). Majority of these dog populations are living in feral condition or semi-independent manner (OrtegaPacheco & Jiménez-Coello, 2011). These unconfined free ranging populations of dogs have unrestricted access to protected areas and to the focused ecosystems where conservation needs are already identified, and that behaviour putting them into proximity with wildlife (Torres & Prado, 2010) which in terms alters natural food chains through direct predation (Newsome et al., 2014). Moreover, dogs take advantages of semi-domesticated life where provisions of leftover or food offered by humans allows them to survive in higher density near or within the protected areas which consequently affect the recovery of natural prey or predator species from fragmented or declining populations (Banks & Bryant, 2007).

India covers approximately 2.3% of the global land area where it holds 23% of the world's carnivore species. These predators still exist with 1.3 billion people, where human density is on average 400 people per square km. Protected reserves spread over about 5.32% (https://wii.gov.in/nwdc_aboutus) of the country's geographical land but many of these areas are not devoid of human presence rather well occupied and exploited for natural resources. Since presence of anthropogenic parameters support population of semi-domestic or feral population, that consequently has impacts on natural predators. In Central India there are four wild canids and striped hyena are the natural predators other than felids, most often found to share lands and other resources with feral dogs.

The golden jackal, grey wolf, striped hyena, and Indian fox play significant roles in human-dominated landscapes of India. They are facing various challenges from feral dogs that range from trophic interaction, food and space competition to altering food habits and disease infestations (Srivathsa et al., 2019; Home et al., 2018; Chourasia et al., 2012; Gherman & Mihalca, 2017).

Conservation efforts for these species require understanding their spatial distribution in and around protected areas or in the landscapes where all these show their presence and surviving from ages. Knowing their ecological coexistence strategy perhaps helps to delineate conservation strategies for the wild canids and hyaenid in and around protected areas.

The Udanti-Sitanadi Tiger Reserve (USTR) in Chhattisgarh, India, is an important habitat for diverse wildlife, including tigers, leopards, and various bird species (Bharos et al., 2018; Basak et al., 2020, 2023). Camera trap surveys have confirmed the presence of rare species like the Indian mouse deer (Basak et al., 2017). However, the reserve faces significant conservation challenges. Low wild prey densities have led to increased livestock predation by tigers and leopards (Basak et al., 2020). Human disturbances and habitat destruction threaten wildlife populations, with many species showing temporal avoidance of human presence (Basak et al., 2023).

To improve understanding overlapping among wild canids and hyaenid with free ranging dogs researchers have employed various methods to establish strong frameworks, that include camera trapping, Relative Abundance Index (RAI) calculations from photo captures, social interviews and occupancy modelling. These all methods in a various ways facilitated to estimate temporal and spatial overlaps (Srivathsha et al, 2019; Bianchi et al, 2020; Malhotra et al, 2021) among these wild and feral species. Consequently, all these methods supported habitat suitability modelling, and assessment of human-wildlife conflicts that feral dogs may cause by increasing competition with wild predators. These studies emphasize the need of such studies to determine the effect of dogs on wild animals and the possible strategies to reduce such menace to protect the biodiversity in USTR and can be repeated as a model elsewhere with same situation.

Joint species distribution models (JSDMs) have emerged as a powerful tool for analyzing community structure and species interactions in ecological research (Poggiato et al., 2022; Zhu et al., 2018). These models extend traditional species distribution models by incorporating correlations among species, potentially improving predictions for rare species and community analyses (Pollock et al., 2014; Wagner et al., 2020). JSDMs can help disentangle environmental effects from biotic interactions, although limitations in separating these factors persist (Poggiato et al., 2021). Applications of JSDMs include informing restoration planning (D'Acunto et al., 2021) and addressing challenges posed by climate-driven species redistribution (Bonebrake et al., 2018). However, the integration of current ecological thinking, including non-equilibrium dynamics, into biodiversity conservation remains a challenge (Wallington et al., 2005). Despite some limitations, JSDMs offer a flexible framework for addressing various questions in fisheries science, ecosystem management, and conservation, potentially improving our understanding of species distributions and community dynamics (Wagner et al., 2020; D'Acunto et al., 2021).

JSDMs utilize statistical and machine learning methods to analyze species co-occurrence patterns and their relationships with environmental variables. Recent advancements include deep learning techniques such as deep neural networks (DNNs), which can capture complex, non-linear interactions and provide more accurate predictions (Poggiato et al., 2021; Hao et al., 2019). These models require robust datasets and significant computational resources for training and validation.

Several studies demonstrate the utility of JDMs in ecological research. Pollock et al. (2014) used JSDMs to study species co-occurrence in Australian forests, revealing significant biotic interactions influencing distributions. Wagner et al. (2020) applied JSDMs to freshwater fish communities, enhancing understanding and prediction of species assemblages. These examples underscore the potential of JDMs to improve ecological insights and support conservation efforts (Pollock et al., 2014; Wagner et al., 2020).

Despite significant advances in species distribution modeling (SDM), current research reveals gaps, particularly in modeling the joint distribution of multiple species. Traditional SDMs often focus on single species, neglecting the interspecies interactions that are critical for accurate ecological predictions (Dormann et al., 2018; Poggiato et al., 2021). These models fail to incorporate complex biotic relationships and environmental dependencies adequately, leading to oversimplified outcomes. Additionally, there is a scarcity of empirical studies applying joint species distribution models (JSDMs) in diverse ecological contexts, limiting the validation and generalization of these models (Zhu et al., 2018; Pollock et al., 2014).

The Udanti-Sitanadi Tiger Reserve (USTR) is home to a variety of canids and hyenids, including the Striped Hyena, Grey Wolf, Golden Jackal, Indian Fox, and free ranging Dogs. These species interact in complex ways, influencing each other's distributions and the overall ecosystem health. There is a pressing need to understand these interactions to inform conservation strategies. Current models are inadequate in capturing the joint distribution and interaction effects among these species. This study aims to address this gap by employing deep neural networks (DNNs) to develop JSDMs that integrate biotic and abiotic factors, offering a more nuanced understanding of species distributions in USTR. Given the limitations of existing models and the critical conservation needs, this study is both timely and necessary (Poggiato et al., 2022; Hao et al., 2019).

The primary objective of this study is to develop and validate JSDMs using deep neural networks to analyze the distribution patterns of canids and hyenids in USTR. Specifically, the study aims to:

1. Map the joint distribution of the Striped Hyena, Grey Wolf, Golden Jackal, Indian Fox, with free ranging Dogs.

2. Identify and analyze the abiotic factors influencing their joint-distribution.

3. Evaluate the effectiveness of DNN-based JSDMs.

This research encompasses the entire USTR, focusing on canids and hyenids due to their ecological significance and diverse presence in the reserve. The study aims to provide comprehensive insights into the joint distribution of these species. However, there are limitations, including the availability and quality of environmental and species interaction data and potential computational constraints associated with complex DNN models.

**Study Area:**

The Udanti-Sitanadi Tiger Reserve (USTR), located in the Gariyaband and Dhamtari districts of Chhattisgarh, central India, spans 1,842.54 km² (Figure 1). It includes the Udanti and Sitanadi Wildlife Sanctuaries as core areas, with the Taurenga, Indagaon, and Kulhadighat Ranges serving as buffer zones. The reserve features a mix of hill ranges and plains within the Mahandi River basin and predominantly comprises of dry tropical peninsular sal forest and southern tropical dry deciduous mixed forest (Champion & Seth, 1968). Dominant vegetation includes Sal (*Shorea robusta*), along with *Terminalia*, *Anogeissus*, *Pterocarpus*, and various bamboo species. The Sitanadi region additionally hosts dry teak forests and mixed deciduous forests, with natural teak patches along alluvial soils near streams and rivers. Riverine species in the area include *Schleichera oleosa*, *Terminalia arjuna*, *Mangifera indica*, *Syzygium cumini*, and several *Ficus sp*.

USTR is a critical habitat for diverse wildlife, including apex predators such as the Tiger (*Panthera tigris*) and co-predators like the Leopard (*Panthera pardus*), Dhole (*Cuon alpinus*), Indian Grey Wolf (*Canis lupus pallipes*), Striped Hyena (*Hyaena hyaena*), and Sloth Bear (*Melursus ursinus*). The reserve supports a range of wild ungulates, providing a substantial prey base, from small species like the Indian Mouse Deer (*Moschiola indica*) and Four-horned Antelope (*Tetracerus quadricornis*) to large species such as the Gaur (*Bos gaurus*) and Sambar (*Rusa unicolor*). Although, Spotted Deer (*Axis axis*) and Wild Pig (*Sus scrofa)* are found to be the main prey species available in the tiger reserve. Smaller carnivores include the Jungle Cat (*Felis chaus*), Rusty-spotted Cat (*Prionailurus rubiginosus*), Golden Jackal (*Canis aureus*), Indian Fox etc. USTR connects with the Sonabeda Wildlife Sanctuary in Odisha, forming the Udanti-Sitanadi-Sonabeda Landscape, which is part of the larger Chhattisgarh-Odisha Tiger Conservation Unit. However, USTR faces significant conservation challenges, including low wild prey densities leading to increased livestock predation, human disturbances, and habitat destruction, causing many species to temporally avoid human presence (Basak et al., 2017; Basak et al., 2020, 2023). Additionally, the landscape is identified as one of the Red Corridors in the country (Putul, 2021) that potentially hindered effective conservation efforts in the whole USTR-Sonabeda landscape.

**Data Collection and Pre-processing**

Species-occurrence data was collected from camera trapping surveys conducted in USTR during 2016-2017 and 2018. The first session, part of the Phase IV tiger monitoring framework, involved 136 camera trap stations in 2x2 km grids across North Udanti, South Udanti, and Kulhadighat ranges. In 2018, under the All India Tiger Estimation (AITE) program, 182 cameras were deployed in 2 sq km grids covering Arsikanhar, Risgaon, Sitanadi, and Kulhadighat ranges. Cameras were placed 4-5 meters from forest trails at knee height to capture clear images of wildlife. Each camera was operational for 30 days per block, totaling 90 days per session. Species presence and absence were recorded, assigning values of 1 and 0, respectively.

GIS analysis was utilized to create a 100-meter grid for finer spatial resolution, incorporating a range of environmental and geographical attributes. Distance variables, including the minimum distance to rivers, roads, and villages, were calculated using GIS techniques in R. Aspect and slope categories, derived from Copernicus 30m Digital Elevation Models (DEM), included north, east, south, and west-facing slopes, as well as various slope steepness categories. Land Use and Land Cover (LULC) categories from ESRI and bioclimatic variables from WorldClim, such as temperature and precipitation metrics, provided a comprehensive climatic profile of the study area. To enhance model performance, input variables were normalized using a quantile transform, facilitating efficient and accurate predictive modeling. Table 1 summarizes the datasets used in this study.

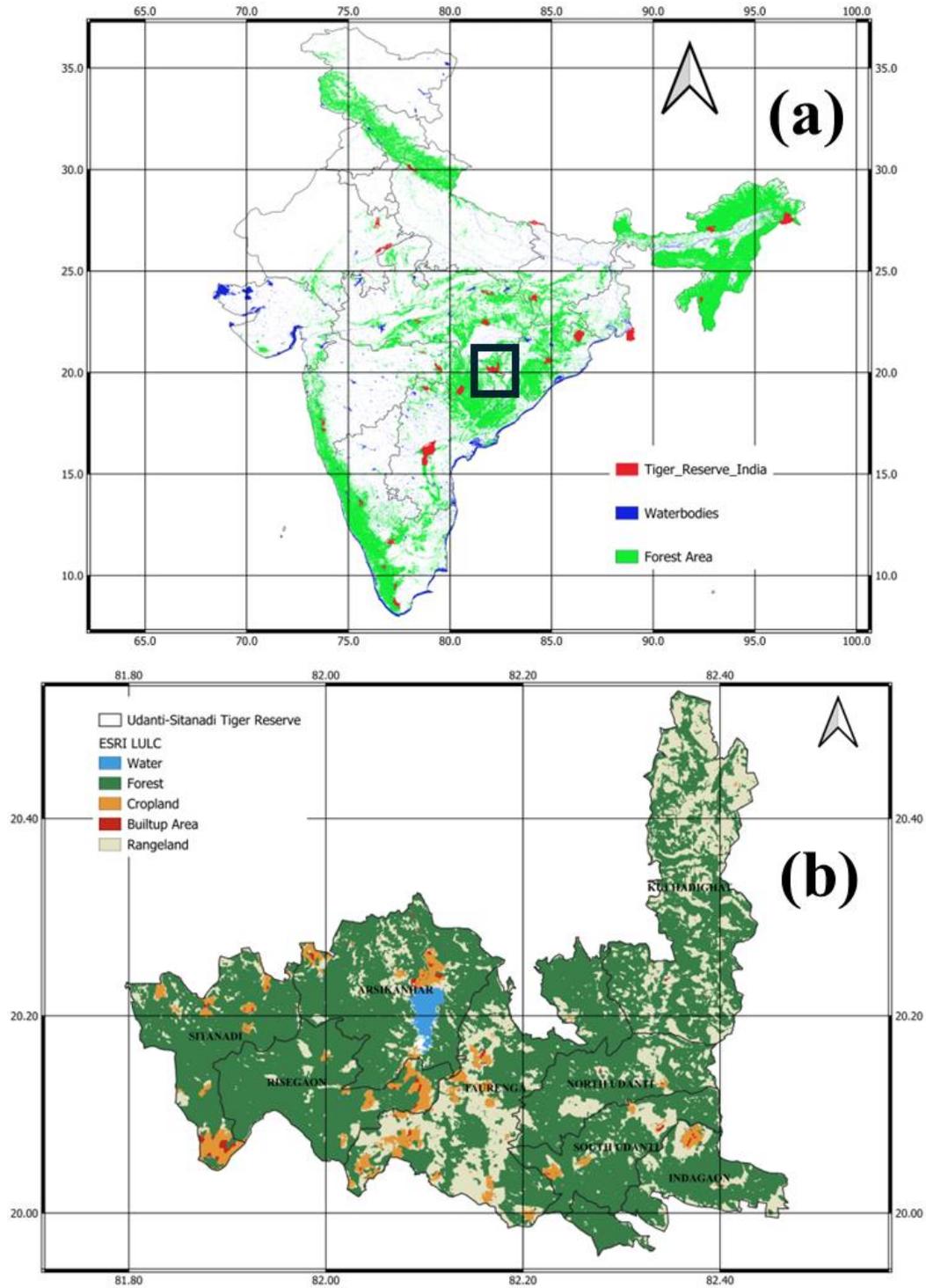

**Figure. 1. (a) Map of Indian landscape complex (CILC) showing location of present study area Udanti-Sitanadi Tiger Reserve (USTR) (a). USTR with the land use/land cover map from ESRI (b)**

**Table 1: The 36 environmental factors used in the construction of the Joint DeepSDM (including bioclimate, human disturbance, topography and vegetation).**

| Short Name | Description | Source |
|---|---|---|
| River_Distances | Distance to nearest river within a 0.1 km grid | OSM, GIS |
| Road_Distances | Distance to nearest road within a 0.1 km grid | |
| Village_Distances | Distance to nearest village within a 0.1 km grid | |
| ASPECT_cat_1 | Fraction of aspect category 1 (e.g., North-facing slopes) | DEM, GIS |
| ASPECT_cat_2 | Fraction of aspect category 2 (e.g., East-facing slopes) | |
| ASPECT_cat_3 | Fraction of aspect category 3 (e.g., South-facing slopes) | |
| ASPECT_cat_4 | Fraction of aspect category 4 (e.g., West-facing slopes) | |
| DEM_COP30 | Digital Elevation Model (DEM) from COPERNICUS | DEM |
| ESRI_LULC_cat_1 | Land Use Land Cover category 1 (water bodies) | ESRI LULC |
| ESRI_LULC_cat_2 | Land Use Land Cover category 2 (rangelands/cropland) | |
| ESRI_LULC_cat_5 | Land Use Land Cover category 5 (built-up areas) | |
| ESRI_LULC_cat_7 | Land Use Land Cover category 7 (shrubland) | |
| ESRI_LULC_cat_11 | Land Use Land Cover category 11 (forests) | |
| SLOPE_cat_1 | Slope category 1 (e.g., gentle slopes, <5 degree) | DEM, GIS |
| SLOPE_cat_2 | Slope category 2 (e.g., moderate slopes, 5-15 degree) | |
| SLOPE_cat_3 | Slope category 3 (e.g., steep slopes, 15-30 degree) | |
| SLOPE_cat_4 | Slope category 4 (e.g., very steep slopes, >30 degree) | |
| wc2.1_30s_bio_1 | Bioclimatic variable 1 (Annual Mean Temperature) | WorldClim Data |
| wc2.1_30s_bio_10 | Bioclimatic variable 10 (Mean Temperature of Warmest Quarter) | |
| wc2.1_30s_bio_11 | Bioclimatic variable 11 (Mean Temperature of Coldest Quarter) | |
| wc2.1_30s_bio_12 | Bioclimatic variable 12 (Annual Precipitation) | |
| wc2.1_30s_bio_13 | Bioclimatic variable 13 (Precipitation of Wettest Month) | |
| wc2.1_30s_bio_14 | Bioclimatic variable 14 (Precipitation of Driest Month) | |
| wc2.1_30s_bio_15 | Bioclimatic variable 15 (Precipitation Seasonality) | |
| wc2.1_30s_bio_16 | Bioclimatic variable 16 (Precipitation of Wettest Quarter) | |
| wc2.1_30s_bio_17 | Bioclimatic variable 17 (Precipitation of Driest Quarter) | |
| wc2.1_30s_bio_18 | Bioclimatic variable 18 (Precipitation of Warmest Quarter) | |

| wc2.1_30s_bio_2 | Bioclimatic variable 2 (Mean Diurnal Range) | |
|---|---|---|
| wc2.1_30s_bio_3 | Bioclimatic variable 3 (Isothermality) | |
| wc2.1_30s_bio_4 | Bioclimatic variable 4 (Temperature Seasonality) | |
| wc2.1_30s_bio_5 | Bioclimatic variable 5 (Max Temperature of Warmest Month) | |
| wc2.1_30s_bio_6 | Bioclimatic variable 6 (Min Temperature of Coldest Month) | |
| wc2.1_30s_bio_7 | Bioclimatic variable 7 (Temperature Annual Range) | |
| wc2.1_30s_bio_8 | Bioclimatic variable 8 (Mean Temperature of Wettest Quarter) | |

## Methodology

A fully connected neural network (Figure 2a) was developed for species distribution modeling, leveraging the PyTorch Lightning framework for efficient training and scalability (Falcon et al., 2019). The model architecture comprised an input layer, one hidden layer, and an output layer:

Input Layer: The input layer size corresponded to the number of abiotic raster layers (predictor variables). These variables are quantile transformed at the data pre-processing stage.

Hidden Layer: A single hidden layer with 31 neurons, activated using the ReLU function to introduce non-linearity. This hidden layer acts as an embedding layer, capturing complex interactions between the input features and allowing the model to learn joint representations of the environmental variables. This is particularly significant in joint species distribution modeling, as it facilitates the understanding of how different environmental factors simultaneously affect multiple species. The choice of 31 neurons was determined through Bayesian hyperparameter optimization, although the detailed results of this optimization process are not included in this paper.

Output Layer: The output layer contained neurons equal to the number of target species, activated using the sigmoid function to generate probabilities for each species' presence. The multi-species, multi-output nature of the model allows it to predict the presence of multiple species simultaneously, capturing inter-species correlations and improving the ecological insights derived from the model.

The model was trained using the Adadelta optimizer, chosen to avoid the need for manual selection of a learning rate (Zeiler, 2012). The training was done on the entire dataset simultaneously due to low number of sample points (393), and the maximum number of epochs was 1000. Early stopping was employed to prevent overfitting, with a patience of 50 epochs based on validation loss. The loss function used was Binary Cross-Entropy (BCE), which is appropriate for binary classification tasks and helps in learning the probability distribution of species presence.The dataset was split into training and validation sets using an 80-20 split. The training set was used to update the model weights, while the validation set monitored the model's performance and applied early stopping. Data loading was managed using the PyTorch DataLoader, which facilitated efficient mini-batch training. Furthermore, Custom callbacks were implemented to log training and validation loss. The EarlyStopping callback monitored the validation loss, and the ModelCheckpoint callback saved the best model based on validation loss. A custom LossHistoryLogger callback recorded the loss history for later analysis. The importance of monitoring both training and validation loss is critical in ensuring the model generalizes well to unseen data and does not overfit to the training set.

The hidden layer, or embedding layer, plays a crucial role in the joint modeling of multiple species distributions. By learning a low-dimensional representation of the input features, the model can capture complex, non-linear relationships between the environmental variables and species presence. This joint representation is essential for understanding how different species respond to the same environmental gradients and for identifying common factors that influence their distributions. Moreover, it enables the model to leverage shared information across species, potentially improving predictions for species with limited occurrence data.

**Model Validation and Performance**

Figure 2b illustrates the training and validation loss curves for the neural network used in species distribution modeling. Both curves show a steep decline in the initial epochs, indicating effective learning and significant loss reduction. Beyond 200 epochs, the curves flatten, suggesting convergence with minimal further improvements. The training loss consistently remains lower than the validation loss, indicating good model fitting without significant overfitting. Early stopping helps identify the optimal point to halt training, with minimal and stable validation loss around the 600th epoch.

Post-training, the best model checkpoint was loaded for further evaluation. The optimal threshold for binary classification was determined using the Youden's J statistic from the ROC curve (Youden, 1950). Metrics such as number of captures for the specific species, ROC-AUC, accuracy, precision, recall, and F1-score were computed for each species using these thresholds. Table 2 summarizes these statistics.

**Table 2: The optimal thresholds and performance metrics (ROC AUC, accuracy, precision, recall, and F1 score) for the species distribution model predictions of Striped Hyena, Grey Wolf, Golden Jackal, Indian Fox, and Dogs in the Udanti-Sitanadi Tiger Reserve.**

| Species | No of captures | Optimal Threshold | ROC AUC | Accuracy | Precision | Recall | F1 Score |
|---|---|---|---|---|---|---|---|
| Striped Hyena | 281 | 0.251 | 0.740 | 0.643 | 0.440 | 0.726 | 0.548 |
| Grey Wolf | 107 | 0.087 | 0.862 | 0.692 | 0.240 | 0.902 | 0.379 |
| Golden Jackal | 22 | 0.042 | 0.911 | 0.735 | 0.127 | 0.938 | 0.224 |
| Indian Fox | 9 | 0.013 | 0.924 | 0.748 | 0.039 | 0.800 | 0.075 |
| Dogs | 1635 | 0.673 | 0.826 | 0.758 | 0.890 | 0.722 | 0.797 |

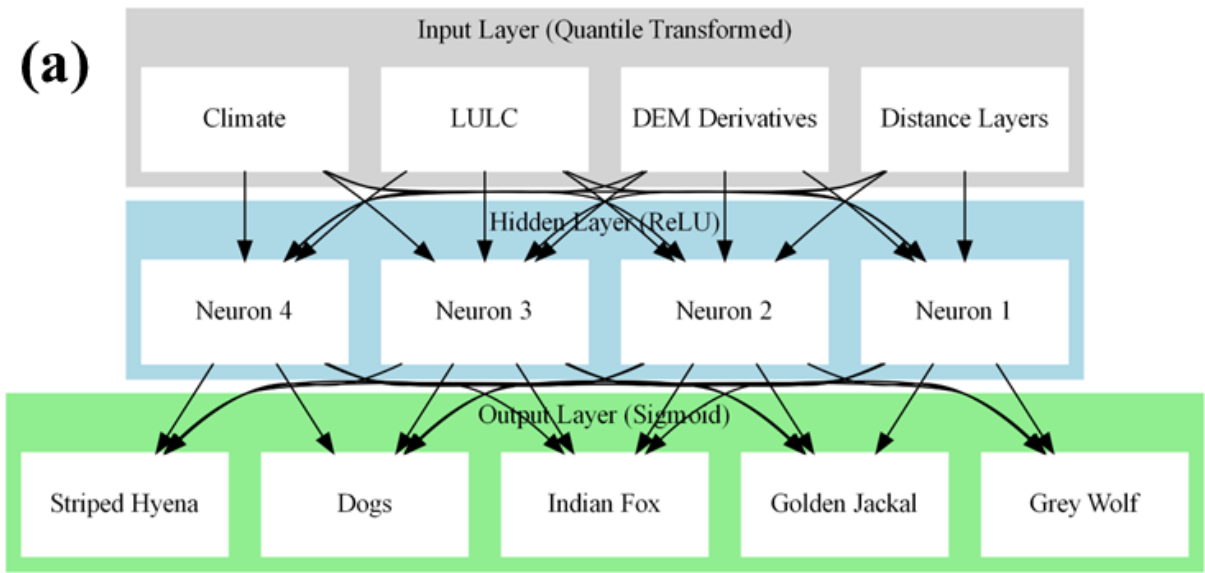

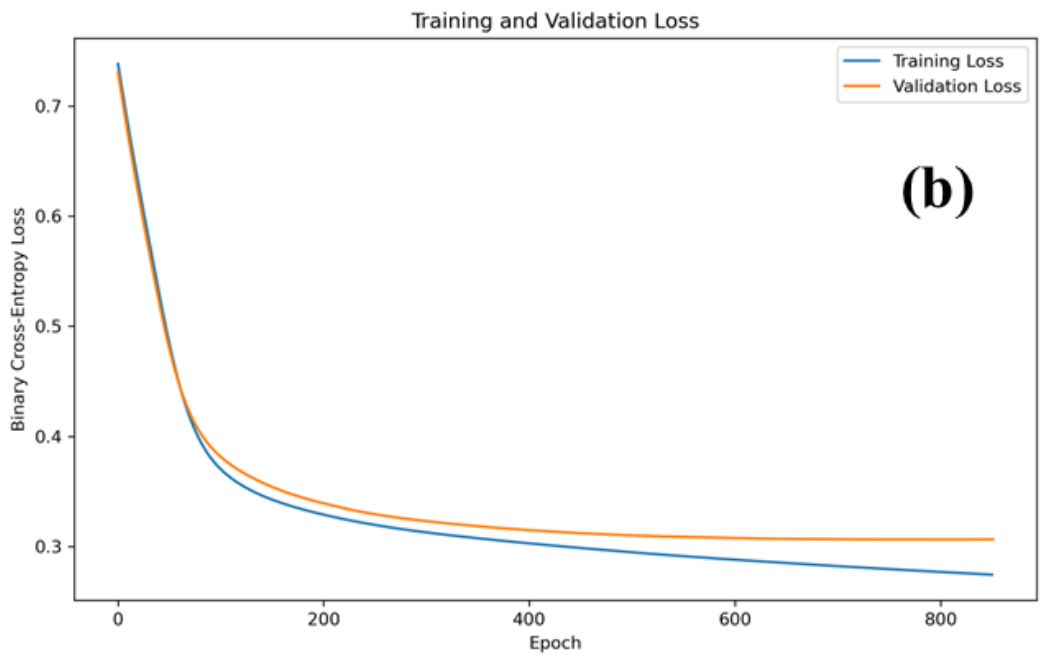

**Figure 2: Neural Network Architecture for Joint Species Distribution Modeling (a), and Model training and validation loss (b)**

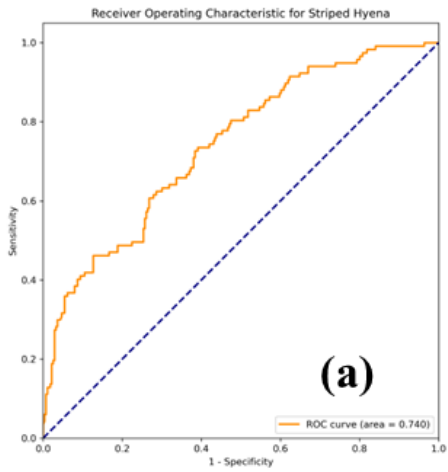
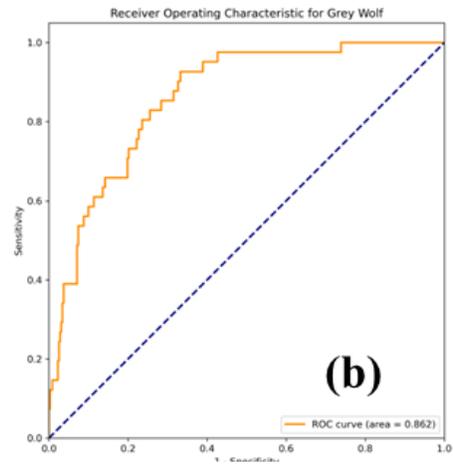
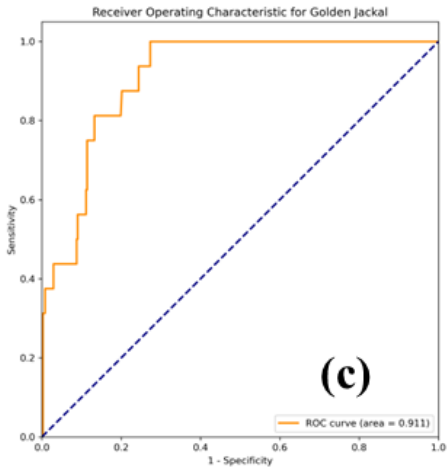
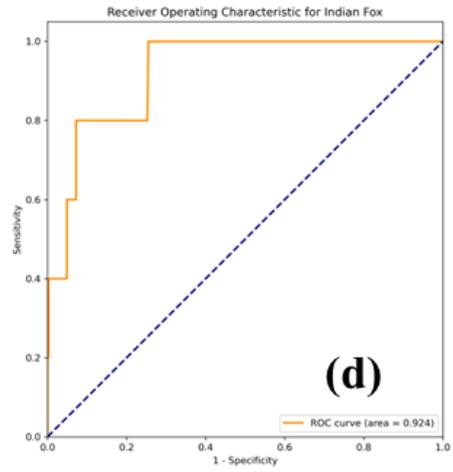
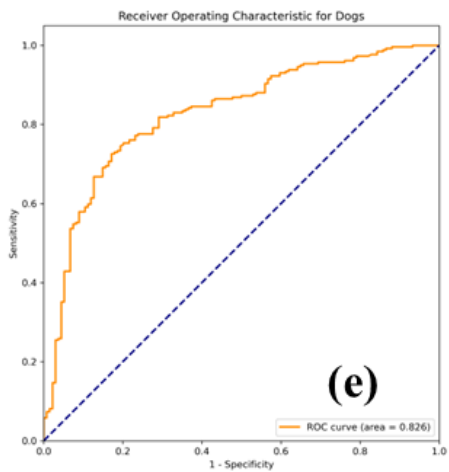

**Figure 3: Receiver Operating Characteristic (ROC) Curves for Dogs (a), Golden Jackal (b), Grey wolf (c), Indian Fox (d), and Striped Hyena (e)**

Figure 3 illustrates the ROC curves for the species distribution models of Dogs, Golden Jackal, Grey Wolf, Indian Fox, and Striped Hyena. The AUC for Dogs is 0.826 (Figure 3a), indicating a strong ability of the model to distinguish between the presence and absence of dogs. For Golden Jackal, the AUC is 0.911 (Figure 3b), reflecting robust model performance. The Grey Wolf model achieves an AUC of 0.862 (Figure 3c), showcasing high discriminative power. The Indian Fox model, with an AUC of 0.924 (Figure 3d), also performs well, indicating a good ability to predict presence and absence. The Striped Hyena model has an AUC of 0.740 (Figure 3e), suggesting moderate discriminative power. Overall, these curves highlight the varying degrees of effectiveness in the models' predictive capabilities, with strong performance for most species and room for improvement for the Striped Hyena model.

**Results**

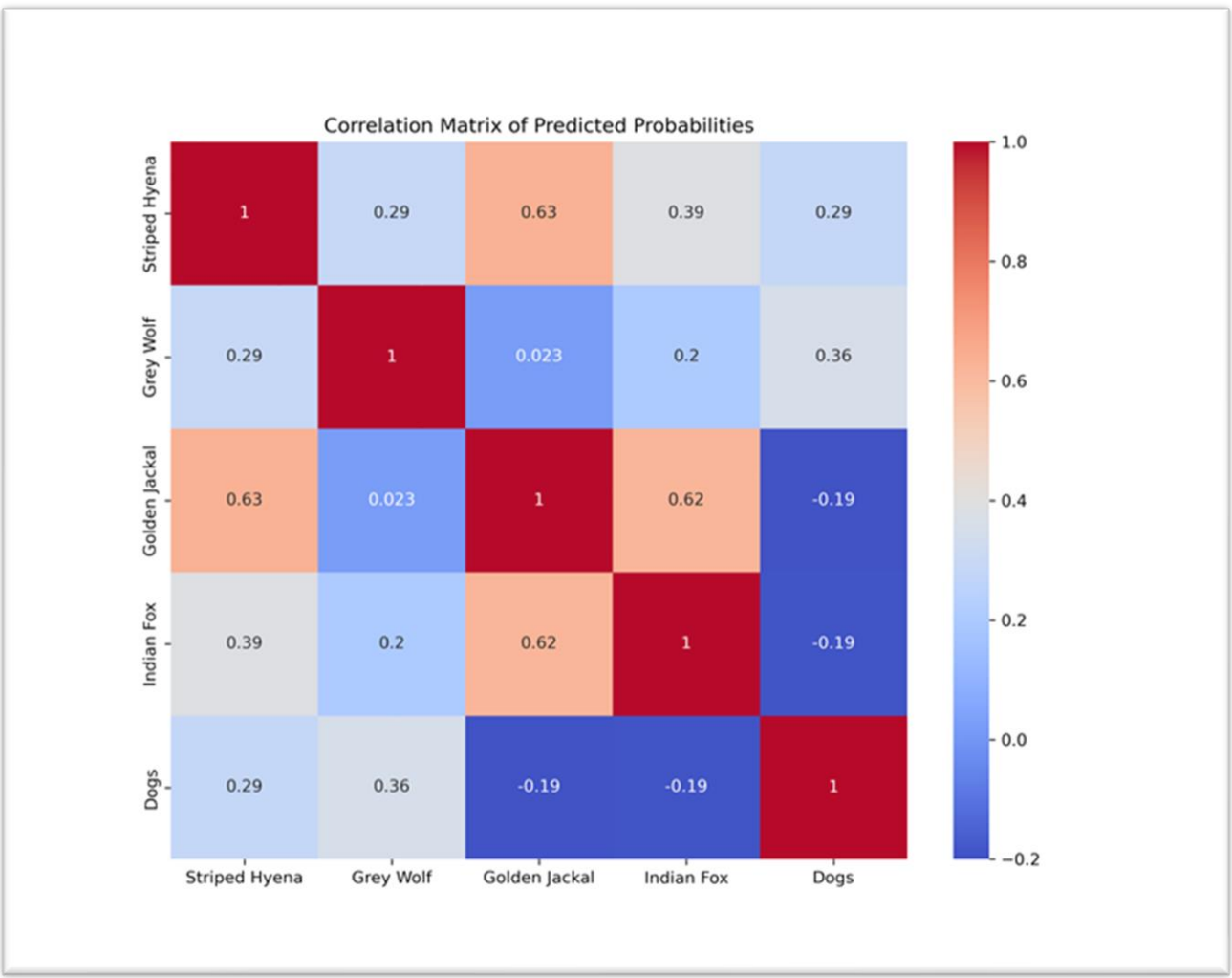

**Figure 4: Correlation Matrix of Predicted Probabilities**

Figure 4 shows the correlation matrix of predicted probabilities for the presence of different species: Striped Hyena, Grey Wolf, Golden Jackal, Indian Fox, and Dogs. The values represent the pairwise Pearson correlation coefficients (p-value<=0.05) between the predicted probabilities for each pair of species. The Golden Jackal shows a moderate to high positive correlation with both the Striped Hyena (0.63) and Indian Fox (0.62), suggesting overlapping habitats or similar environmental preferences. Conversely, the correlation between the Golden Jackal and Grey Wolf is very low (0.023), indicating distinct ecological niches. Dogs exhibit a moderate positive correlation with the Grey Wolf (0.36) and a weaker correlation with the Striped Hyena (0.29), potentially reflecting some space or environmental overlap. However, the correlations between Dogs and both the Golden Jackal and Indian Fox are weak and negative (-0.19 for both), further indicating distinct distribution patterns. These correlations help elucidate the potential habitat overlaps and ecological relationships among these carnivores in the study area.

Figure 5 presents SHAP (SHapley Additive exPlanations) summary plots for five carnivore species in the Udanti-Sitanadi Tiger Reserve (USTR): (a) Striped Hyena, (b) Grey Wolf, (c) Golden Jackal, and (d) Indian Fox. These plots visualize the impact of various environmental variables on the predicted probability of each species' presence, with the x-axis representing the magnitude of this impact. Positive SHAP values indicate that a variable increases the likelihood of species presence, while negative values decrease it. These plots help identify the most influential environmental factors determining the distribution of each species and provide insight into the consistency and variability of these impacts across the study area.

For the Striped Hyena (Figure 5a), gentle slopes (SLOPE_cat_1), high precipitation during the wettest quarter (wc2.1_30s_bio_16), and forested areas (ESRI_LULC_cat_11) are positively associated with their presence, while north-facing slopes (ASPECT_cat_1) and temperature seasonality (wc2.1_30s_bio_15) decrease their likelihood of presence, indicating less favorable conditions. In contrast, the Grey Wolf (Figure 5b) avoids cropland and rangeland areas (ESRI_LULC_cat_2), preferring gentle slopes (SLOPE_cat_1) and areas with higher precipitation, both annually (wc2.1_30s_bio_12) and during the wettest quarter (wc2.1_30s_bio_16), which likely support a stable prey base. The Golden Jackal (Figure 5c) shows a strong preference for west-facing slopes (ASPECT_cat_4) and areas with stable temperature seasonality (wc2.1_30s_bio_15), favoring moderate slopes (SLOPE_cat_2) and high precipitation during the wettest quarter (wc2.1_30s_bio_16), which together create an optimal habitat. The Indian Fox (Figure 5d) prefers west-facing slopes (ASPECT_cat_4) but avoids cropland/rangeland areas (ESRI_LULC_cat_2), suggesting a preference for less disturbed environments. Higher annual precipitation (wc2.1_30s_bio_12) and precipitation during the wettest quarter (wc2.1_30s_bio_16), as well as north-facing slopes (ASPECT_cat_1), are also favorable, reflecting the fox's need for consistent water availability and stable conditions.

These findings underscore the importance of terrain, temperature, and land cover in shaping the distribution patterns of these carnivores within USTR, providing valuable insights for targeted conservation and habitat management strategies. It's important to note that the distribution of domestic dogs is not necessarily dependent on natural habitat characteristics and often dependent on resource provisioning by humans. Their presence is closely tied to human settlements; wherever humans are, dogs are likely to be present, regardless of the natural habitat. This is why we didn't discuss the influence of abiotic factors on the dog's habitat in detail.

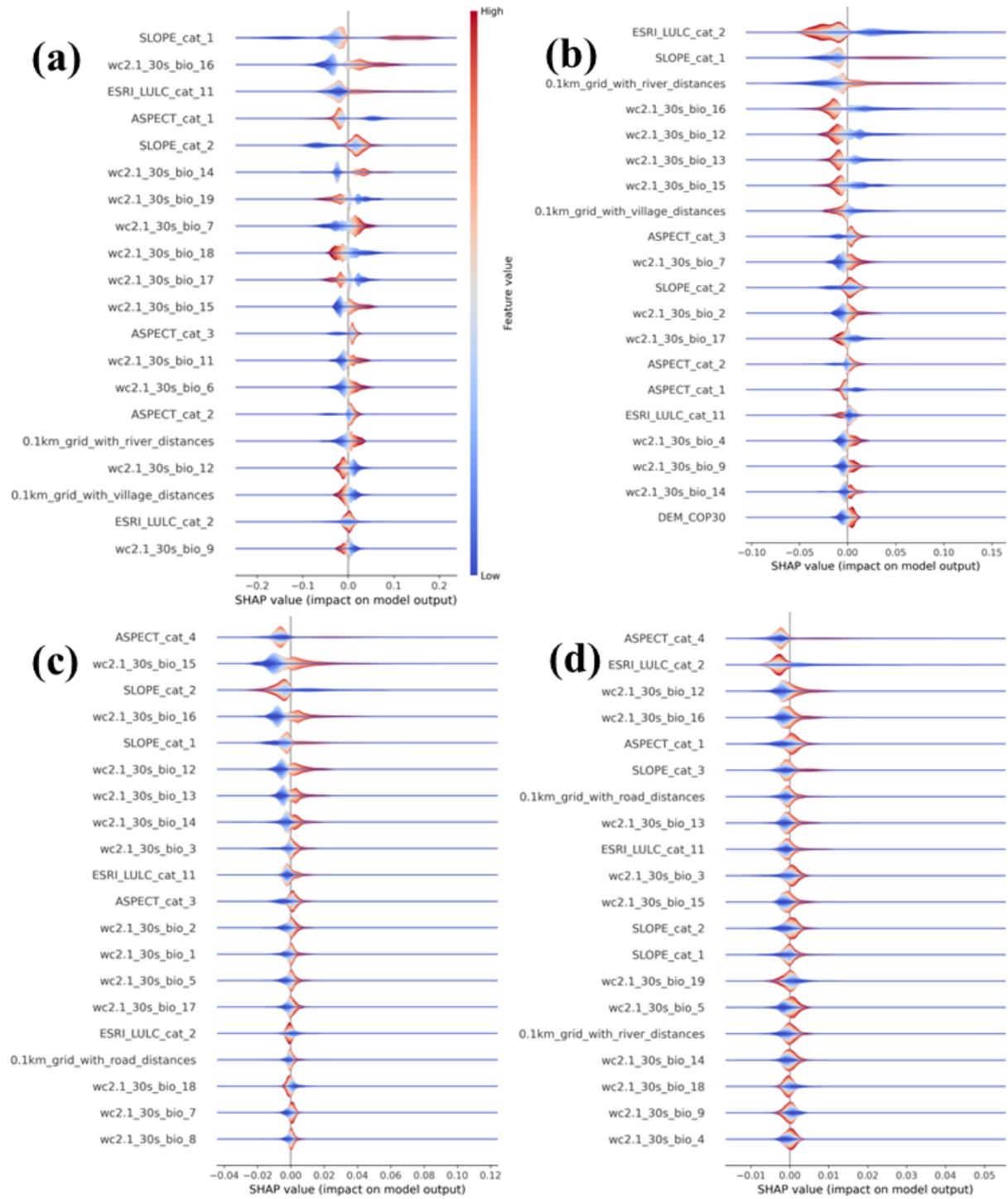

**Figure 5: SHAP Summary Plots of Environmental Variable Importance (a) Striped Hyena, (b) Grey Wolf, (c) Golden Jackal, and (d) Indian Fox in USTR**

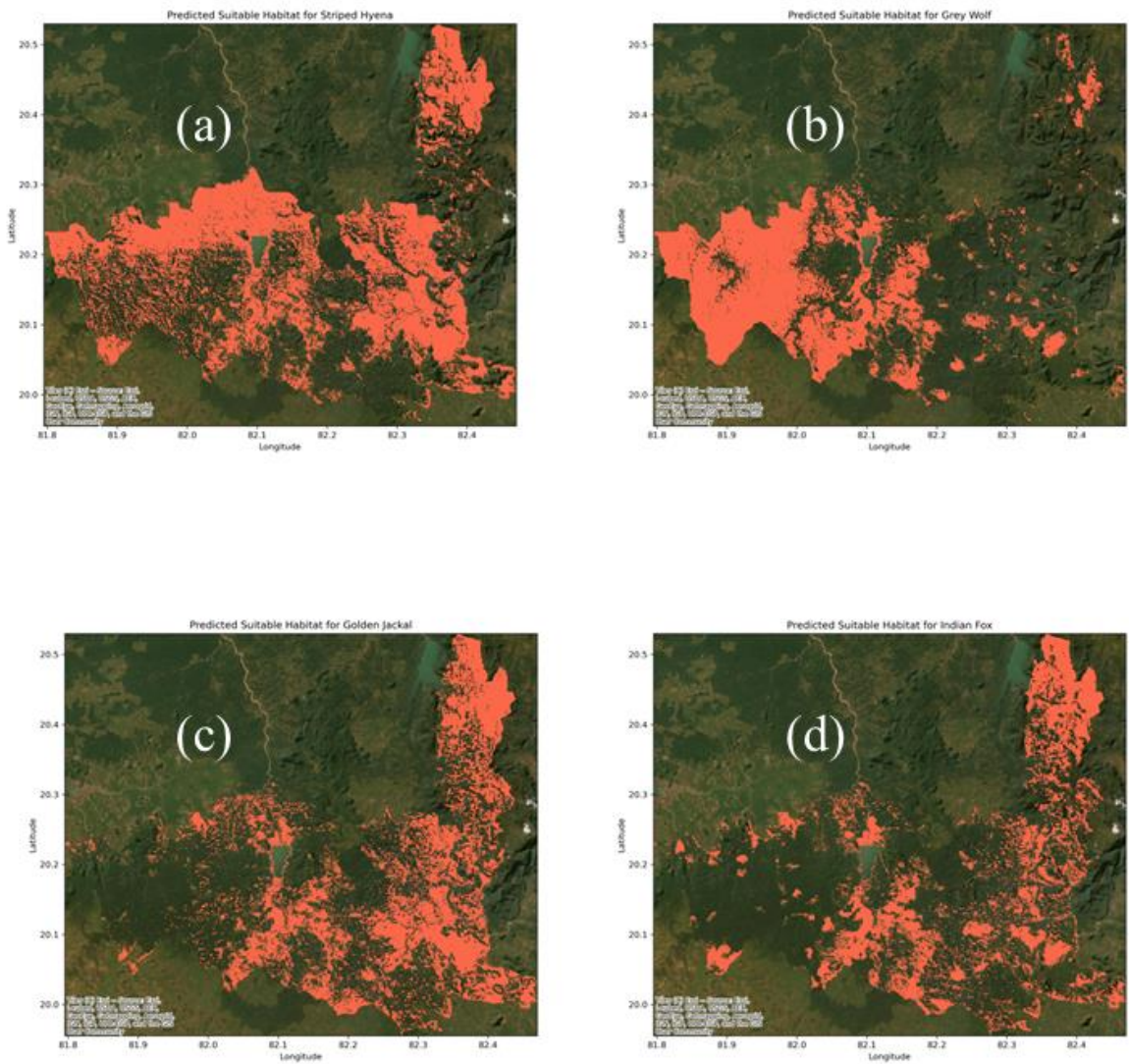

**Figure 6: Habitat Distribution of (a) Striped Hyena, (b) Grey Wolf, (c) Golden Jackal, and (d) Indian Fox in USTR**

Figure 6 displays the predicted suitable habitats for five carnivore species in the Udanti-Sitanadi Tiger Reserve (USTR):(a) Striped Hyena, (b) Grey Wolf, (c) Golden Jackal, and (d) Indian Fox . These habitats were estimated by applying the optimal threshold for each species, determined using the Youden method. The Youden method identifies the threshold that maximizes the sum of sensitivity and specificity, effectively balancing the true positive and true negative rates. This approach ensures that the predicted habitats are as accurate as possible, highlighting areas within the reserve where each species is most likely to thrive.

The Striped Hyena (Figure 6a) occupies a vast, continuous area, particularly in the North Sitanadi and Udanti regions, demonstrating its adaptability to diverse environments, including degraded dry deciduous forests and less forested areas, though some fragmentation appears in the northeastern Kuladighat area. The Grey Wolf (Figure 6b) shows a more concentrated presence in the Sitanadi region, with smaller, fragmented patches in Udanti and Kuladighat, likely reflecting its preference in higher human abundant areas toand rugged terrain in Kuladighat has minimum presence. The Golden Jackal (Figure 6c) stands out for its extensive and continuous habitat across the Udanti and Kuladighat regions, extending into Sitanadi, underscoring its adaptability to various environmental conditions, including human-altered landscapes. In contrast, the Indian Fox (Figure 6d) exhibits a more restricted and fragmented habitat, primarily in the central and eastern Udanti and Kuladighat regions, indicating its reliance on specific, less common microhabitats and its narrower ecological niche compared to the other carnivores in the reserve. The free ranging and semi-domesticated Dogs (Not shown in the figure) are primarily found in the Sitanadi and northeastern Kulhadighat area, with smaller, scattered patches in the central and southeastern parts indicating close relationship to human settlement. The predicted suitable habitats for the carnivores in the Udanti-Sitanadi Tiger Reserve reveal distinct patterns for each species.

These habitat predictions emphasize the distinct ecological niches occupied by carnivore species within the Udanti-Sitanadi Tiger Reserve (USTR). The northeastern region, particularly Kulhadighat, is a crucial area for Indian Foxes, reflecting their preference for the less disturbed rugged terrain, found in this part of the reserve. In contrast, the central and southern regions, including Sitanadi and Udanti, are vital for Striped Hyenas, Grey wolf, and Golden Jackals showcasing their adaptability to more varied and human-influenced environments within USTR.

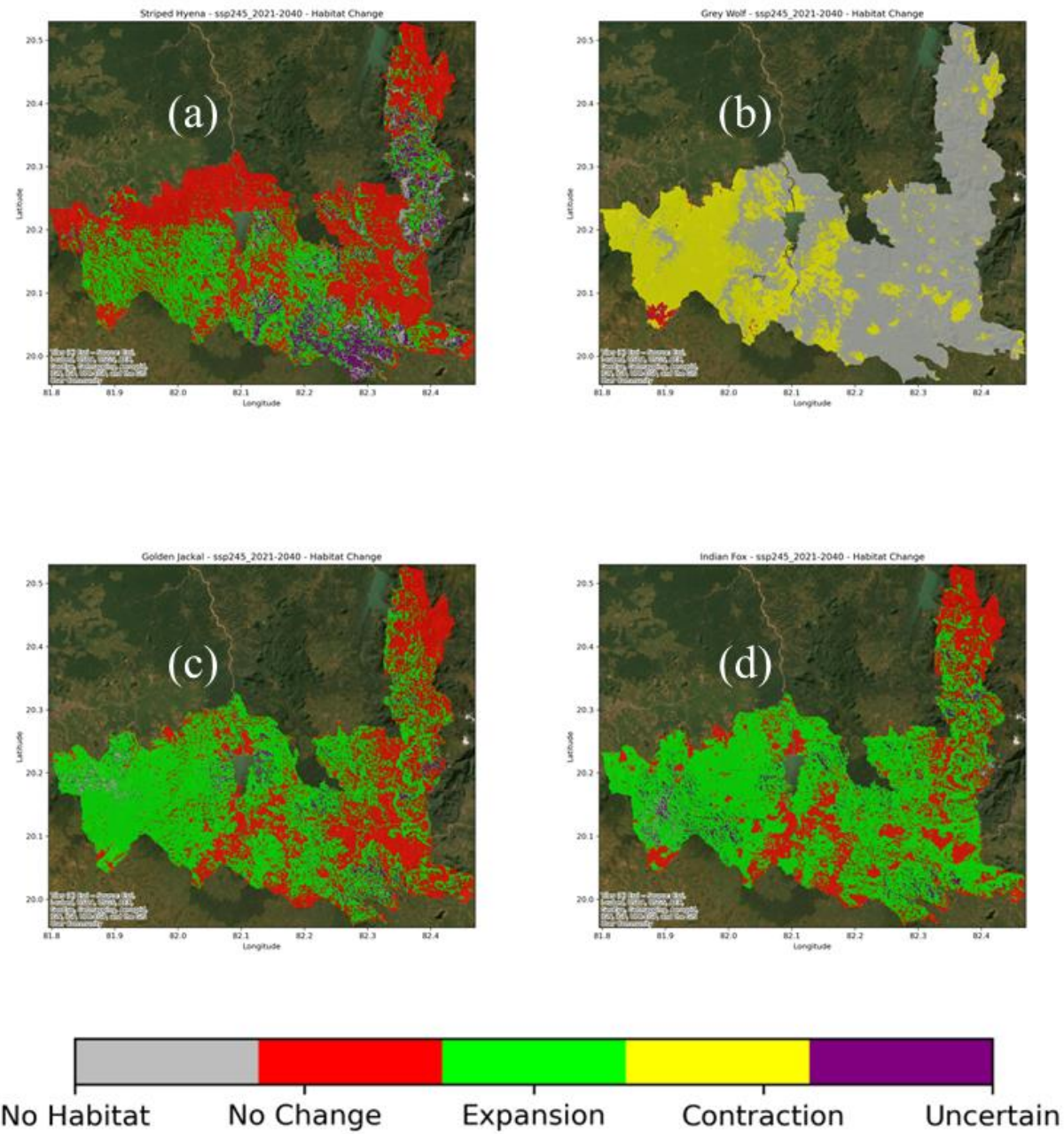

**Figure 7: Predicted Habitat Changes for Striped Hyena (a), Grey Wolf (b), Golden Jackal (c), India Fox (d), (e) in USTR Under Climate Scenario SSP245 (2021-2040)**

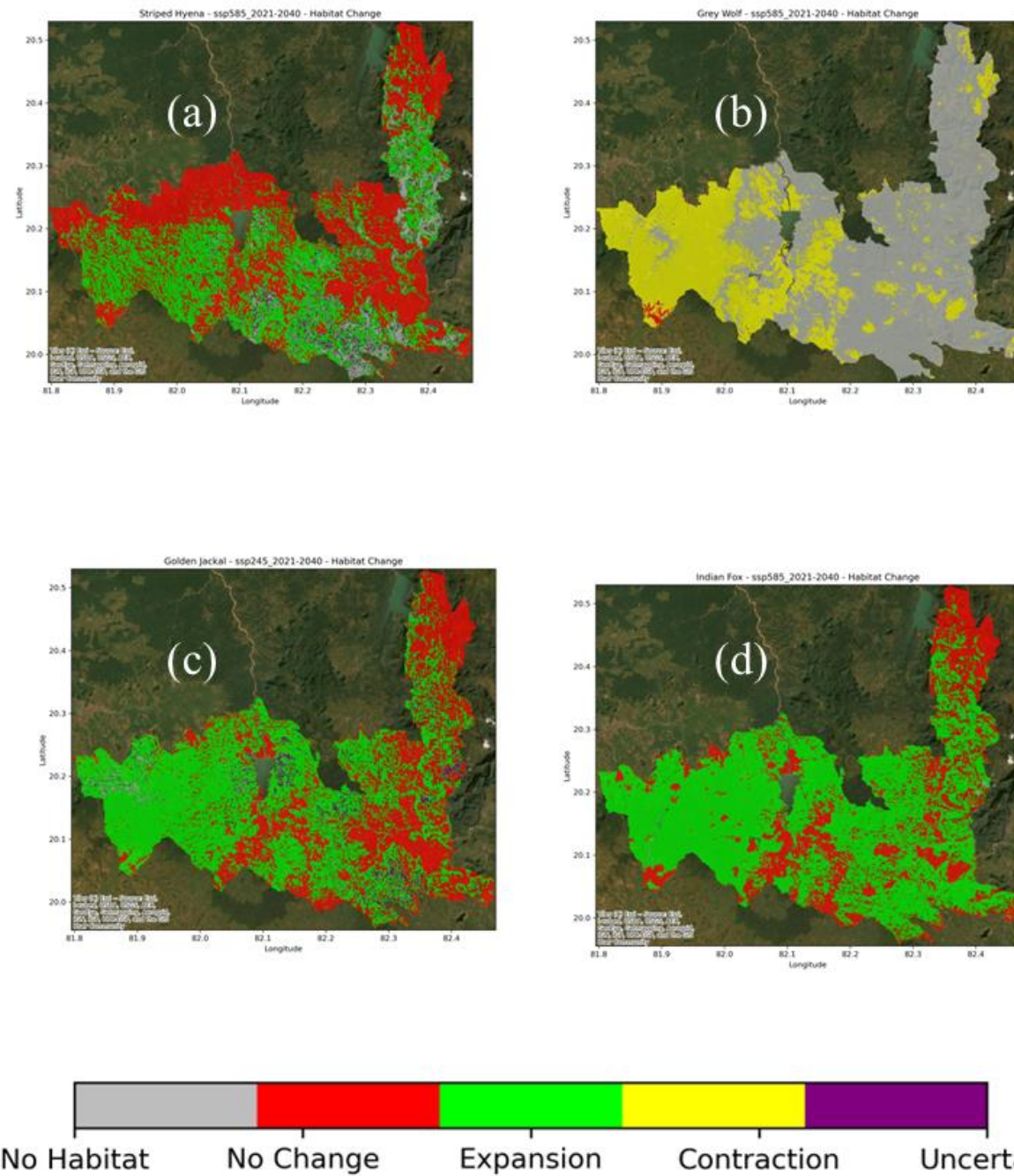

**Figure 8: Predicted Habitat Changes for Striped Hyena (a), Grey Wolf (b), Golden Jackal (c), India Fox (d), (e) in USTR Under Climate Scenario SSP585 (2021-2040)**

**Table 3. The table presents habitat changes across different climate change scenarios for the periods 2021-2040 using four climate models: CMCC-ESM2, GISS-E2, HadGEM3-GC3, and UKSM1 under SSP245 and SSP585 scenarios. The categories include Original Habitat, No Habitat, No Change, Habitat Expansion, Habitat Contraction and Uncertain.**

| Species | Scenario | Original Habitat (sq km.) | No Habitat (sq km.) | No Change (sq km.) | Habitat Expansion (sq km.) | Habitat Contraction (sq km.) | Uncertain (sq km.) |
|---|---|---|---|---|---|---|---|
| Striped Hyena | ssp245_2021-2040 | 972.91 | 76.02 | 958.81 | 676.93 | 0.45 | 227.02 |
| Grey Wolf | ssp245_2021-2040 | 696.99 | 1250.49 | 9.72 | 0 | 674.46 | 4.56 |
| Golden Jackal | ssp245_2021-2040 | 719.88 | 20.51 | 713.84 | 1127.4 | 0 | 77.48 |
| Indian Fox | ssp245_2021-2040 | 571.84 | 13.86 | 566.19 | 1272.53 | 0 | 86.65 |
| Dogs | ssp245_2021-2040 | 727 | 1213.56 | 22.26 | 0 | 651.88 | 51.53 |
| Striped Hyena | ssp585_2021-2040 | 972.91 | 127.69 | 960.54 | 755.08 | 0.08 | 95.84 |
| Grey Wolf | ssp585_2021-2040 | 696.99 | 1250.49 | 5.63 | 0 | 680.29 | 2.82 |
| Golden Jackal | ssp585_2021-2040 | 719.88 | 33.21 | 713.91 | 1170.1 | 0 | 22.01 |
| Indian Fox | ssp585_2021-2040 | 571.84 | 2.58 | 566.19 | 1367.99 | 0 | 2.47 |
| Dogs | ssp585_2021-2040 | 727 | 1213.83 | 12.34 | 0 | 690.25 | 22.81 |

The habitat change predictions for five carnivore species in the Udanti-Sitanadi Tiger Reserve (USTR) were analyzed under two different Shared Socioeconomic Pathways (SSP) climate change scenarios: SSP245 and SSP585, for the period 2021-2040. These scenarios were modeled using four climate models: CMCC-ESM2, GISS-E2, HadGEM3-GC3, and UKSM1. The SSP245 scenario represents an intermediate pathway, where global societal development follows a moderate pace, leading to balanced mitigation and adaptation strategies. In contrast, the SSP585 scenario represents a high-emission, fossil-fuel-intensive future with rapid economic growth and limited mitigation. The predicted habitat changes are depicted in Figures 7 and 8, using a color-coded scheme: green for habitat expansion, red for no change , yellow for contraction , purple for uncertain changes, and gray for areas where no suitable habitat exists. The changes are shown only for locations where the direction or neutrality of the change is agreed upon by at least three out of the four models. Grid points where there is no consensus among the models are marked as uncertain. Table 3 presents the summary of the amount of area showing different types of changes under different climate change scenarios.

Figure 7 presents the predicted habitat changes for four carnivore species in the Udanti-Sitanadi Tiger Reserve under the SSP245 scenario for 2021-2040. The Striped Hyena (Figure 7a) shows mixed habitat expansion and contraction, with significant uncertainty in the northeastern regions. The Grey Wolf (Figure 7b) faces extensive habitat contraction, especially in the central and southern areas, with minimal uncertainty and no expansion. In contrast, the Golden Jackal (Figure 7c) is expected to see considerable habitat expansion, although some areas may experience contraction, with limited uncertainty. The Indian Fox (Figure 7d) also shows broad habitat expansion, particularly in the central regions, but faces potential contraction in the northeast, with notable uncertainty in some areas. Overall, these predictions highlight the varied responses of these species to future climate and land-use changes, emphasizing the need for adaptable conservation strategies. Figure 8 illustrates the predicted habitat changes for four carnivore species in USTR under the SSP585 scenario for 2021-2040. The results are almost same as the scenario SSP245 with varying degree of uncertainty in certain areas.

Comparing the results from SSP245 and SSP585, it becomes clear that the high-emission scenario (SSP585) leads to more pronounced and extensive habitat changes for all species. The Grey Wolf (Figures 7b and 8b) experiences significant habitat contraction under both scenarios, but the loss is more severe under SSP585, with very limited areas of stability or expansion. The Golden Jackal (Figures 7c and 8c) demonstrates remarkable adaptability, showing substantial habitat expansion in both scenarios; however, the area of contraction increases under SSP585, particularly in the northeastern regions. The Indian Fox (Figures 7d and 8d) generally benefits from habitat expansion in both scenarios, but SSP585 introduces greater uncertainty and some contraction, indicating potential challenges in certain areas. The Striped Hyena (Figures 7a and 8a), which shows a mix of expansion and stability under SSP245, faces more severe habitat contraction under SSP585, particularly in the northeastern part of USTR, along with increased areas of uncertainty. These findings highlight the diverse and species-specific impacts of climate change on carnivore habitats within USTR, emphasizing the need for adaptive and scenario-based conservation strategies to mitigate these effects.

**Discussion**

The application of Joint Species Distribution Models (JSDMs) combined with deep neural networks provided a powerful tool for understanding these dynamics, allowing for the simultaneous consideration of multiple species and their interactions with the environment. These approaches are increasingly essential for studying large carnivores, which face significant threats such as habitat loss, hunting, and prey depletion, leading to population declines and range contractions (Karanth & Chellam, 2009; Ripple et al. 2014). Worldwide 80 different wild species are affected by free ranging dogs and 31 among them are listed in IUCN threatened category. Even the carnivores are not spared from their attacks and often faced competitive harassments (Homme et al., 2019). Therefore, in this current research our emphasis was on checking joint distribution of all the wild canids and hyena with free ranging or free ranging dogs in USTR. We found significant role of free ranging dogs in shaping carnivore distributions in the tiger reserve. Sitanadi, being a more human-populated region, where presence of human associated species like free ranging or free ranging dogs pose distinct challenges for wild canids of all sizes and striped hyenas. This approach provided a more accurate and nuanced understanding of species distributions in USTR, offering valuable insights for conservation planning (Poggiato et al., 2022; Pollock et al., 2014).

Despite having a status of a tiger reserve USTR is a prey depleted landscape and occupied by 99 villages across the reserve, where natural resource extractions including hunting of wildlife for local consumption by communities are in regular practice. Furthermore, these villages hold large livestock population with density of 14.48/km2 (Basak et al., 2020). Habitats altered by humans and by providing food to dogs gave their population a set of opportunities to thrive. Dogs' access to food and habitats that have been modified by humans have allowed their numbers to flourish. Dogs were traditionally kept in villages to protect their owners from stray animals and other potential hazards, but they were also frequently utilized as hunting companions. Dogs are usually seen exploring the forest in the absence of people and to kill wild ungulates and other species whenever they had the opportunity, albeit this study lacks the chance to evaluate the quantity of wild prey is utilized by dogs in USTR, but they do offer similar scopes for more research in the tiger reserve.

In India's agro-pastoral environments, wolves and striped hyenas are recognized as the top predators (Nayak et al., 2015). Previous studies (Nayak et al., 2015; Maurya et al., 2017; Majgaonkar et al., 2018; Saren et al., 2019) have reported the co-occurrence of these wild predators in Rajasthan, Bihar, Maharashtra, Jharkhand, and West Bengal. The current study revealed that grey wolves and striped hyenas are widely distributed in the USTR's human-dominated areas, which are also home to a large number of semi-domestic or feral dogs. Smaller canids, like foxes and jackals, have slightly negative correlations where dogs are more prevalent, according to camera trap captures. These species show adaptability to human-modified environments, which is reflected in their widespread distribution in this area. In contrast, species like the Indian Fox, which prefer less disturbed habitats, are more concentrated in the mountainous, less populated Kuladighat region. This difference underscores the importance of considering human impact when planning conservation strategies, as species have varying levels of tolerance to human presence and habitat alteration (Van der Weyde et al., 2021).

This research highlights how specific environmental factors, such as slope, temperature, and precipitation, play a crucial role in shaping the distributions of carnivores within the Udanti-Sitanadi Tiger Reserve. The SHAP summary plots reveal that species like the Golden Jackal and Indian Fox are particularly influenced by these variables, with the Jackal favoring west-facing slopes and stable temperature conditions, and the Fox preferring areas with consistent precipitation and less rugged terrain. Similarly, the Striped Hyena shows a strong association with gentle slopes and forested areas, while the Grey Wolf avoids cropland, preferring higher precipitation areas that likely support a stable prey base. These insights are vital for understanding the ecological niches of these species and predicting their responses to future environmental changes, informing more targeted and effective conservation strategies in the reserve (Siepielski et al., 2018).

The study also highlights the potential impact of climate change on carnivore habitats in USTR. Under both SSP245 (a moderate climate change scenario) and SSP585 (a high-emission scenario), significant habitat shifts are predicted for the five studied species. For instance, the Grey Wolves are likely to experience substantial habitat contraction in both scenarios, which suggests that their current habitats in human-modified landscapes might not remain viable under more extreme future conditions. The Striped Hyena, although currently widespread, also shows a potential decrease in suitable habitat under these scenarios, particularly in areas with higher human activity like Sitanadi. These findings indicate that climate change will likely exacerbate the existing pressures on these species, necessitating more adaptive and forward-thinking conservation approaches (Abade et al., 2014; Smith et al., 2017).

In summary, this research underscores the need for integrated conservation strategies that consider both scope of intensive studies on impacts of dogs on local wildlife, and future climate change. Species in USTR are likely to face significant challenges from dogs due to 1) spatial overlapping, 2) food competition, 3) predation 4) fear driven behavioural changes 5) disease transmission and 6) hybridization (Ritchie et al., 2014; Wierzbowska et al., 2016; Banks & Bryant, 2007; Silva-Rodríguez & Sieving, 2012; Zapata-Ríos & Branch, 2016; Vanak et al., 2014; Acosta-Jamett et al., 2011; Knobel et al., 2014; Leonard et al., 2014). Finding such challenges of exclusive agents like dogs on wildlife highlights the importance of introducing steps to regulate such adverse effects with strict management interventions possible to use in the landscape. Moreover, the results signify preserving less disturbed areas like Kuladighat, which provide critical habitats for more sensitive species. The use of advanced modeling techniques, such as deep neural networks within JSDMs, offers a promising path forward for predicting and managing the impacts of global change on biodiversity (Wagner et al., 2020).

**Conclusion and Future Directions**

The free ranging or semi-domesticated dog populations in and around forests are primarily sustained by human resource supply. These dog populations can have a number of detrimental effects on domestic wildlife, and human health. In addition to their supremacy in competition, they harbor deadly viruses that seriously endanger wildlife populations (Doherty et al., 2017).

Dogs were found to be largely present towards Sitanadi region of USTR where densely located settlements are well observed, possibly as a result of higher food supply from local communities. We found large spatial overlapping of dogs with canids and striped hyenas which may have adverse impacts on wild predators in the long run. Therefore, we recommend interventions by means of capturing and translocation of dogs, vaccinations to control disease transmission, sterilization to control hybridization and birth control of free ranging or semi-domesticated dogs as mentioned in Standard Operating Procedure (SOP) to deal with stray/feral dogs in Tiger Reserve, complied by National Tiger Conservation Authority (NTCA). There are obvious involvements of communities in these procedures because for them, dogs are trustworthy companions with intense bonding, on that account prior information of such planning, awareness of the reason of such interventions and inclusion of communities during the implementation are most important to mitigate the issues.

There are many tiger reserves in India are facing the same consequences, and it is crucial for park authorities to introduce various invasive strategies to control the free ranging dog populations. We recommend through our present study, how concerned authorities, research workers and implementing Non-Governmental Organizations can use JSDM by combining deep neural network to strongly assess joint distribution of different carnivores with dogs. The predicted habitat shifts under different climate scenarios further highlight the need for adaptive conservation strategies that can mitigate the anticipated impacts of global change on biodiversity in USTR.

Building on the findings of this study, future research should focus on several following key areas;

1) First, incorporating more detailed temporal data and finer-scale environmental variables could improve the accuracy of species distribution models, especially under varying climate change scenarios.

2) Additionally, exploring more advanced neural network architectures, including ensemble methods and temporal dynamics, would provide a deeper understanding of how species interactions and environmental changes influence distribution patterns over time.

3) Further studies should also examine the effects of habitat connectivity and fragmentation, particularly in human-modified landscapes like Sitanadi to develop more targeted conservation strategies.

4) In the face of climate change, the predicted habitat extinction for the Grey Wolf in the Udanti-Sitanadi Tiger Reserve underscores the urgency of adaptive conservation strategies. Under the SSP245 and SSP585 scenarios, the Grey Wolf may lose its viable habitat entirely, particularly in areas with higher human activity. This highlights the critical need for conservation efforts that not only address current human impacts but also proactively mitigate the future challenges posed by climate change. Future research should focus on integrating climate projections into species distribution models to better inform conservation planning and ensure the survival of this species in a rapidly changing environment.

5) Finally, integrating socio-economic factors into ecological models could offer a more holistic approach to conservation planning, ensuring that the needs of both wildlife and local communities are addressed in efforts to preserve precious biodiversity in USTR.

In situations when adverse parameters are governing the ecology of wild carnivores in human-modified landscapes, this entire method has the potential to be applied globally to develop effective carnivore conservation plans that may be implemented inside or outside of protected areas (PAs).

## Acknowledgment

We express earnest gratitude to Shri Kaushalendra Singh, former PCCF (Wildlife) and Shri Sudir Agarwal, PCCF (Wildlife), Chhattisgarh. We also convey our gratitude to Mr. K. Murugan, former CCF (Wildlife) for his initiative and continuous support during the project implementation period. We would like to convey our deepest gratitude to Shri O. P. Yadav, Member Secretary Chhattisgarh for his trust in Nova Nature Welfare Society and provided with every possible support. We are always grateful to Chhattisgarh State Forest Department to keep faith on us and providing necessary permission and essential financial support to conduct the study. We are thankful to Shri B.V. Reddy, Divisional Forest Officer, Shri Nair Vishnu Narendran, Divisional Forest Officer and Shri Ayush Jain, Divisional Forest Officer for their constant support during the field. We also want to thank Shri Sunil Sharma, Sub- Divisional Forest Officer for his directions in the field, without such guidance it might be impossible to collect data from the difficult terrain of Udanti Sitanadi Tiger Reserve. We would like to convey our sincere thanks to biologist Mr. Chiranjivi Sinha for his rigorous contribution in the field during the tiger monitoring program. We also thank Mr. Ajaz Ahmed, Mr. Nitesh Sahu, Mr. Om Prakash Nagesh and the entire team from Nova Nature Welfare Society for their contribution to field works, and all frontline forest staff from USTR for their support during the study.